\documentclass{article}
\usepackage{authblk}
\usepackage[left=2cm,right=2cm,top=2.25cm,bottom=2.25cm,headheight=12pt,letterpaper]{geometry}
\usepackage[utf8]{inputenc}
\usepackage[T1]{fontenc}
\usepackage{amsfonts}
\usepackage{braket,array}
\usepackage{multirow}
\usepackage{xcolor}
\usepackage{booktabs}
\usepackage{comment}
\usepackage{graphicx}% Include figure files
\usepackage{dcolumn}% Align table columns on decimal point
\usepackage{bm}% bold math
\usepackage{amsmath}
\usepackage{xr,color}
\usepackage[square,sort,comma,numbers]{natbib}
\usepackage[superscript,biblabel,nomove]{cite}
\usepackage{mathptmx,soul}
\newcommand{\pare}[1]{\left( #1 \right)}

\usepackage[labelfont={bf,sf},labelsep=period,justification=raggedright]{caption}
\usepackage[colorlinks,citecolor=red,urlcolor=blue,bookmarks=true,hypertexnames=true]{hyperref}

\date{}
\title{\huge \textbf{Generation of heralded vector-polarized single photons in remotely controlled topological classes}}

\author[1,+]{Samuel Corona-Aquino}
\author[1,+]{Zeferino Ibarra-Borja}
\author[2]{Omar Calderón-Losada}
\author[3]{Bruno Piccirillo}
\author[3]{Verónica Vicuña-Hernández}
\author[1]{Tonatiuh Moctezuma-Quistian}
\author[4]{Dorilian Lopez-Mago}
\author[1]{Héctor Cruz-Ramírez}
\author[1,*]{Alfred B. U'Ren}
\affil[1]{Instituto de Ciencias Nucleares, Universidad Nacional
Autónoma de México, Apartado Postal 70-543, Ciudad de México 04510, México.}
\affil[2]{Centro de Investigacion e Innovación en Bioinformatica y Fotonica,  Edificio E20
No.~1069, Universidad del Valle, Cali, Valle del Cauca, 760042, Colombia}
\affil[3]{Dipartimento di Fisica “Ettore Pancini”, Università degli Studi di Napoli Federico II, Napoli, 80126, Italy}
\affil[4]{Escuela de Ingeniería y Ciencias, Tecnologico de Monterrey, Monterrey, N.L. 64849, Mexico}

\affil[*]{alfred.uren@correo.nucleares.unam.mx}

\affil[+]{these authors contributed equally to this work}

\begin{document}
\maketitle

\begin{abstract}
    We demonstrate an experimental protocol for the preparation and control of heralded single photons in inhomogeneously polarized states, such as Vector Vortex and Full Poincar\'e  beam states. A laser beam is shaped by a voltage-controlled spin-to-orbital angular momentum converter $q$-plate device which eliminates the need for an interferometer for the robust preparation of high-quality inhomogeneously polarized beams. Such a beam is then used as pump in a spontaneous parametric downconversion (SPDC) photon-pair source. We demonstrate the full pump to heralded single photon transfer of the intensity/phase distributions, as well as of the vector polarization structure. Additionally, we show that by controlling the polarization to which the heralding idler photon is projected before detection, we can toggle between the direct and basis-switched pump-single photon transfer.  We show that this non-local control of the heralded single photon pertains also to the topological class of the resulting heralded single photon.  We believe that our work will lead to new opportunities in photons-based quantum information processing science.
\end{abstract}

\section*{Introduction}
Polarization singularities of an optical mode are points on the wavefront with an undefined local orientation of the polarization ellipse, surrounded by a region of high phase gradient~\cite{bernabeu_phase_2020}.  Intimately related to phase singularities,  polarization singularities are encountered in beams with an inhomogeneous polarization distribution~\cite{RoadmapRubinsztein,Shen2024}; in the paraxial regime, they can be viewed as a superposition of two orthogonally polarized states with a spatially varying phase difference. Specifically, the superposition of two modes with opposite helicity (with topological charges $\ell$ and $-\ell$), and circularly polarized with opposite handedness results in a vector vortex (VV) mode. The resulting polarization is inhomogeneously distributed, locally linear except on the optical axis, at which the orientation is undefined, creating a V-point singularity~\cite{Darvehi_2021}. As a second relevant example, the superposition of a circularly polarized, Gaussian TEM$_{00}$ mode ($\ell=0$) with a helical mode with circular polarization of the opposite handedness results in a mode well described by the stereographic projection of the Poincar\'e sphere; the specific projection is determined by the relative phase of the superposition.  Such beams are typically known as  Full Poincar\'e (FP) beams~\cite{FPB_Alonso}, and the optical field over their transverse wavefronts is mainly elliptical with points of undefined orientation known as C-points~\cite{Darvehi_2021}. 

In the last two decades, the unique optical properties of polarization singularities -- that stem from the structural inseparability of  Orbital Angular Momentum (OAM) and Spin Angular Momentum (SAM)  --  have been exploited for an increasing number of applications in both the classical and quantum domains. In the classical domain, VV beams that obey axial amplitude/phase symmetry are sometimes referred to as Cylindrical Vector beams~\cite{Zhan:09}.  In particular, VV beams with radial polarization have attracted interest because they provide optimal focusing, leading to a strong and localized longitudinal field component~\cite{TightFocus,Dorn2003} which has enabled many applications in imaging, including confocal microscopy, two-photon microscopy, second-harmonic generation microscopy~\cite{Biss:03}, third-harmonic generation microscopy~\cite{Carrasco:06}, and dark field imaging~\cite{Carrasco:06}.  This longitudinal component  has been shown to permit stable 3D trapping of metallic nanoparticles~\cite{Tweezers}, empower laser micromachining,~\cite{V_G_Niziev_1999,meier_material_2007}, implement finer optical lithography, increase packing density for optical storage~\cite{Lithography}, and enable single molecule spectroscopy~\cite{Spectroscopy}. 

Moving on to FP beams, they have been used to unveil topological properties of light~\cite{Rosales-Guzman_2018,Forbes2021}, particularly for optical Möbius strips and optical Skyrmions~\cite{Bauer2015, Gutierrez-Cuevas_2021}. They have also shown potential for polarimetry, involving the use of the resulting polarization pattern for single-shot measurements~\cite{SuarezBermejo2019}, and for beam shaping using, e.g. resulting in flattop focusing~\cite{Han2011a}. 

In the quantum domain, specifically for photon pairs generated by the spontaneous parametric downconversion (SPDC) process, vector polarization properties open up new avenues for the engineering of 
correlations and entanglement involving the polarization and OAM degrees of freedom.   For example, entangled pairs of VV photons~\cite{PhysRevA.94.030304} of arbitrary order have been generated using a $q$-plate, a birefringent retardation waveplate with a topologically charged optic-axis distribution that can couple or decouple spin and orbital angular momentum distributions ~\cite{piccirillo_orbital_2013}. 
Hyperentanglement between time-frequency modes and VV modes has been also demonstrated~\cite {PhysRevResearch.2.043350}, thus enabling protocols such as complete Bell-state analysis~\cite{PhysRevA.68.042313} and logic gate slimming~\cite{lanyon_simplifying_2009}. 
WhilevVV beams have been generated both actively~\cite{pohl_operation_2003,naidoo_controlled_2016} -- i.e. via laser intracavity devices -- and passively -- through extra-cavity devices~\cite{Rubano:19} -- entangled photons in VV modes have so far been generated  in a two-step sequence.   Thus, single photons are first generated in standard modes, which are subsequently transformed into structured modes using dedicated optical components~\cite{FicklerPol,PRA10Hybrid,PhysRevA.94.030304,PhysRevResearch.2.043350,10.1117/1.AP.5.4.046008}. However, generating structured single photons directly at the source could prove to be more convenient for  compactness,  flexibility and/or scalability in the implementation of quantum information protocols. The polarization and phase structure of VV beams has been transferred from the pump to the signal and idler in Stimulated Parametric Down-Conversion (StimPDC)~\cite{PhysRevApplied.15.024039}. In Spontaneous Parametric Down-Conversion (SPDC), however, to our best knowledge, only the transfer of the scalar vortex structure from the pump to single photons has been already demonstrated~\cite{Vero16,Jabir,Forbes2021Basis}. 
In this study, we demonstrate the generation of \emph{vector-polarized} single photons directly from the SPDC process. Our experimental design incorporates a voltage-controlled $q$ plate for the robust and alignment-free preparation of VV or FP beams, which are used to pump a polarization-entangled photon pair source based on a dual type I SPDC crystal~\cite{Kwiat1999}, resulting in the generation of hybrid-entangled photon pairs. As will be seen, in such a source, projecting the polarization of one of the photons of a given pair before its detection allows non-local control of the vector mode of its twin (including its topological class), paving the way for proposing new quantum information protocols that exploit this capability.
\section*{Theory}
A $q$-plate is an electro-optical geometric phase element consisting of a liquid crystal based retardation waveplate with a topologically charged axis distribution, capable of converting photon spin angular momentum (SAM) into orbital angular momentum (OAM) and vice versa. The action of a $q$-plate can be described by the  operator shown in Equation~(\ref{Eq:SVAP}), where the state $\ket{e,\ell}$ represents an optical mode with polarization $e\in\{L,R\}$ and OAM topological charge $\ell=2q$ (in  a quantum-inspired ket notation)~\cite{Darvehi_2021}. 
In equation \ref{Eq:SVAP}, $\phi$ is the spatial azimuthal polar angle and $\alpha_0$ a constant angle representing a uniform azimuthal shift with respect to the frame of reference. The Pancharatnam-Berry phase $\pm 2\pare{   q \, \phi + \alpha_0}$, imparted to the input photons, when $\delta\neq 2 k \pi$,  has a helical structure with topological charge $2q$, and is accompanied by a polarization handedness reversion. The retardation $\delta$ is electrically controlled using an applied external voltage of a few volts. When $\delta=(2 k+1)\pi$, where $k$ can be any integer number (half-wave operation), the $q$-plate achieves maximum conversion efficiency and the polarization handedness is fully reversed. 
\begin{center}
\begin{equation}
 \hat{Q}\pare{\delta,q} = \cos\frac{\delta}{2}\pare{\ket{L, 0}\bra{L, 0} + \ket{R, 0}\bra{R, 0}} 
 + i\sin \frac{\delta}{2}\left(e^{i 2 \alpha_0}e^{i 2 q\phi}\ket{R, 2q}\bra{L, 0}  + e^{-i 2 \alpha_0}e^{-i 2 q\phi}\ket{L, -2q}\bra{R, 0} \right).
 \label{Eq:SVAP}
\end{equation}
\end{center}
When the input polarization state is linear and $\delta=(2 k+1)\pi$, a V-point is formed. In contrast, when input polarization is circular and $\delta=(2 k+1)\pi/2$, a C-point is formed. Specifically, when $\delta=\pi/2$ and the impinging beam is a TEM$_{00}$ mode in the polarization state $\ket{L}$, the output beam becomes a coherent superposition of a right-circularly polarized beam with geometric phase $\pm (2q\phi+\alpha_0)$ and a left-circularly polarized TEM$_{00}$ mode. If the impinging light has $\ket{R}$ polarization, the output beam becomes a coherent superposition with handedness and geometric phase reversed (compared to the aforementioned case). For $\delta=\pi$, a linearly polarized TEM$_{00}$ mode is turned into a  coherent superposition of right- and left-handed polarized beams with opposite topological charge. Thus, by selecting the phase retardation $\delta$  and the input polarization state, we can select the superposition, leading to either VV beams or FP beams. 

Therefore, our $q$-plate enables the preparation of the pump for the SPDC process in the form
\begin{equation}
\ket{\psi}_\mathrm{pump} = \frac{1}{\sqrt{2}}\pare{\ket{L,\ell_{p}} + e^{i\varphi} \ket{R,\ell_{p}'}},
    \label{Eq:VecPump_Circular}
\end{equation}
where $\varphi$ is a fixed phase that can be controlled by a half-wave plate following the $q$-plate. Note that such a state includes VV beams and FP beams as special cases, namely, $\{\ell_{p},\ell_{p}'\}=\{j,-j\}$ for VV beams and $\{\ell_{p},\ell_{p}'\}=\{j,0\}$ for FP beams, where $j\in \mathbb{N}$.

In the following section, we discuss our experiments using a type-I dual-crystal SPDC source. 
%The crystals are pumped by a laser beam with a spatial structure, such as the one in Eq.~\ref{Eq:VecPump_Circular}. 
In the first (second) crystal the process $H
\rightarrow V V$ ($V \rightarrow H H$) occurs, which involves the annihilation of a horizontally- (vertically-) polarized pump photon and the generation of vertically (horizontally) polarized photon pairs.  The resulting SPDC state can be expressed as
\begin{multline}\label{Eq:State}
\ket{\psi}_\mathrm{SPDC} =\frac{1}{2}\sum_{\ell_i, \ell_i',\ell_i'',\ell_i'''}
\left[c_{\ell_i}\ket{V,\ell_{p}-\ell_{i}}_s\ket{V,\ell_{i}}_i + e^{i\varphi}c_{\ell_i'}\ket{V,\ell_{p}'-\ell_{i}'}_s\ket{V,\ell_{i}'}_i \right.\\
\left. - i e^{i \phi} \pare{c_{\ell_i''}\ket{H,\ell_{p}-\ell_{i}''}\ket{H,\ell_{i}''}_i - e^{i\varphi}c_{\ell_i'''}\ket{H,\ell_{p}'-\ell_{i}'''}_s\ket{H,\ell_{i}'''}_i}\right],
\end{multline}   
\noindent where the subscripts refer to the signal (s) and idler (i) photons and $\phi$ represents a possible phase difference in the contributions to the state from the two crystals.
Since we are interested in the preparation of the signal photon state, we project the two degrees of freedom of the idler photon, i.e. OAM and polarization, in a specific way before its detection. First, we project the SPDC two-photon state in such a way that only the vanishing OAM contribution remains for the idler photon. This is achieved experimentally by coupling the idler photon to an optical fiber, resulting in the state 
\begin{equation}
\ket{\Psi} = \vert  \ell=0 \rangle_i \mbox{}_i\langle \ell=0 \vert \psi \rangle_\mathrm{SPDC}=
\frac{1}{4}\big[ \pare{\ket{V,\ell_{p}}_{s} + e^{i\varphi} \ket{V,\ell_{p}'}_{s}}\ket{V,0}_{i} 
- ie^{i\phi} \pare{\ket{H,\ell_{p}}_s - e^{i\varphi} \ket{H,\ell_{p}'}_s}\ket{H,0}_{i}\big].
\label{SPDC_prep}
\end{equation}
Second, we additionally project the former two-photon state onto each of the four idler-photon polarizations $\vert A \rangle_i$, $\vert D \rangle_i$, $\vert L \rangle_i$, and $\vert R \rangle_i$, resulting in the state $\ket{\Psi} \rightarrow \vert e \rangle_i \mbox{}_i\langle e \vert \Psi \rangle \equiv \vert u_e \rangle_s \vert e,0 \rangle_i$, where $\vert u_e \rangle_s$ represents the conditioned signal-photon state in the following manner
\begin{align}
  \ket{u_A}_s & = \frac{1}{\sqrt{2}}\pare{\ket{L,\ell_{p}}_s + e^{i\varphi^\prime} \ket{R,\ell_{p}'}_s}\label{Eq:AProj}  \\
  \ket{u_D}_s & =  \frac{1}{\sqrt{2}}\pare{\ket{R,\ell_{p}}_s + e^{i\varphi^\prime} \ket{L,\ell_{p}'}_s} \label{Eq:DProj}  \\
  \ket{u_L}_s & = \frac{1}{\sqrt{2}}\pare{\ket{D,\ell_{p}}_s + e^{i\varphi^\prime} \ket{A,\ell_{p}'}_s}  \label{Eq:LProj} \\
  \ket{u_R}_s & = \frac{1}{\sqrt{2}}\pare{\ket{A,\ell_{p}}_s + e^{i\varphi^\prime} \ket{D,\ell_{p}'}_s}   \label{Eq:RProj}
\end{align}
where $\varphi' = \varphi+\pi$. Here we have assumed that there is no phase difference between the contributions from the two crystals, i.e. $\phi=0$. 
Note from equations (\ref{Eq:VecPump_Circular}) and (\ref{Eq:AProj}) that the vector-conditioned structure of the signal photon resulting from the projection of the idler photon to the state $\vert A,0 \rangle_i$ is the same as that of the pump except for a $\pi$ phase shift in $\varphi$. 
As will be seen in both experiments and simulations, this $\pi$ phase shift results in a rigid rotation of the transverse intensity and position-dependent polarimetry by an angle depending on the characteristic disclination index $I_C$ of the pump polarization structure. Specifically, the latter angle is $\pi/N$, where $N=|2(I_C-1)|$, for $I_C\neq 1$, is the number of radial lines characterizing the polarization topography~\cite{Vicuña-Hernández_2023}. Thus, for this particular idler-photon projection, the full amplitude of the pump vector (including transverse intensity, phase, and polarization) is transferred to the signal photon within a rigid rotation depending on $I_C$.

It is interesting to consider the effects of projecting the idler photon to each of the remaining three states, i.e., $\vert D,0 \rangle_i$, $\vert L,0 \rangle_i$, and $\vert R,0 \rangle_i$.  Note that when projecting to the state $\vert D,0 \rangle_i$, the roles of the polarizations $L$ and $R$ are reversed with respect to the state $\ket{u_A}_s$.  Projecting to the state $\vert L,0 \rangle_i$ has the effect of changing the basis, from $\{L,R\}$ to $\{D,A\}$, with respect to the projection to $\vert A,0 \rangle_i$.  Finally, note that similarly to the transformation observed when changing the projection state from $\vert A,0 \rangle_i$ to $\vert D,0 \rangle_i$, when changing the projection state from $\vert A,0 \rangle_i$ to $\vert D,0 \rangle_i$, the roles of the $D$ and $A$ polarizations are reversed.  Thus, due to the additional richness provided by optical fields exhibiting vector polarization, we are not only able to transfer the amplitude from the pump to a conditioned single photon (as was done previously in ref.~\cite{Sibeiro,Vero16}), but we have the additional ability to transform the transferred amplitude according to the change of basis and/or the $L$/$R$ or $D/A$ permutations described above. This is important because it leads to an additional degree of control over the heralded single-photon; for example, in the FP case, it allows us to obtain an FP heralded single photon in different topological classes (i.e., lemon vs. star) simply by changing the idler photon projection for the very same setup; this is in contrast to previous work that requires a different setup for each topological class~\cite{Khajavi_2016}.
\begin{figure*}[h!]
    \centering
    \includegraphics[width = 0.9\textwidth]{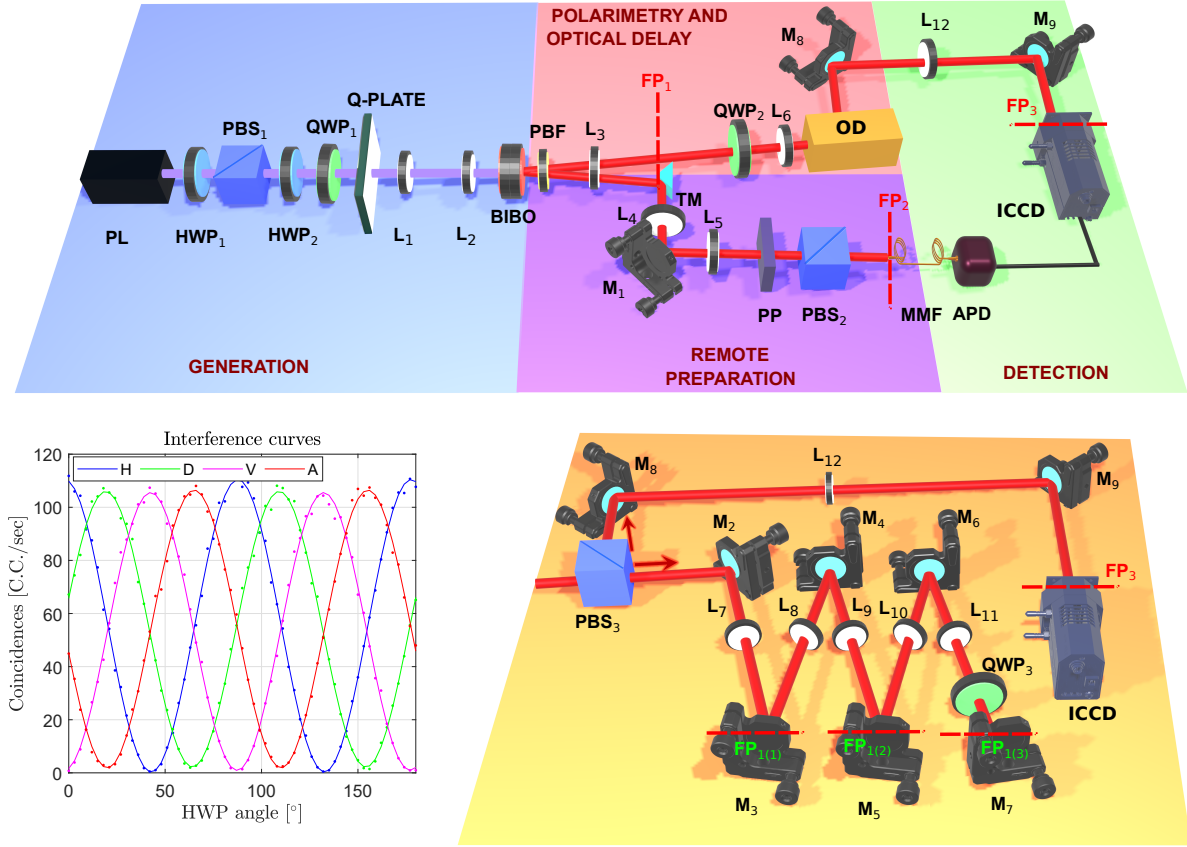}
    \caption{Experimental setup used for heralding of single photons with controllable FP or VV polarization patterns.  A $q$-plate device is used to prepare the pump in a high quality vector mode (Generation).  While the idler photon (Remote preparation) is projected to a specific polarization and zero OAM prior to its detection, the signal photon (polarimetry and optical delay) is transmitted through an image-preserving delay line (see bottom right for detail),  and is sent to a polarimeter prior to detection by an ICCD camera (Detection).  Our polarization correlation measurements are presented in the bottom left panel.
    }
    \label{fig:setup}
\end{figure*}
\section*{Experiment}

We have designed an experimental setup, see Figure ~\ref{fig:setup}, in which non-collinear, polarization-entangled SPDC photon pairs are generated in a dual Bismuth Borate (BiBO) crystal~\cite{Rangarajan2009}, followed by the projection of the idler photon to its zero OAM contribution $\ell=0$, as well as to each of the four polarizations $\vert A \rangle_i$, $\vert D \rangle_i$, $\vert L \rangle_i$, $\vert R \rangle_i$ in turn.  The detection of the projected idler photon then heralds the signal photon, which is detected in coincidence, in a spatially and polarization-resolved manner~\cite{Delay2}, relying on a polarimeter followed by an ICCD camera.    

The pump for the SPDC process is a narrow-band diode laser (Moglabs ECD004; DL) operating at $\lambda_p=405$ nm and structured with a voltage-controlled $q$-plate device.  A half-wave plate (HWP$_1$) followed by a polarizing beam splitter (PBS$_1$) optimizes the resulting horizontally polarized laser flux. The combination of a quarter wave plate (QWP$_1$) followed by one of three $q$-plate devices characterized by their $q$ values: 1/2, 1, or 3/2, is then used to structure the pump into a high quality VV beam or FP beam. The type of pump beam produced is determined by: i) the choice of $q$-plate device, ii) the orientation angle of QWP$_1$ to obtain either linear or circular incident polarization, and iii) the voltage applied to the $q$-plate. For example, to generate FP pump beams, we rely on the quarter-wave mode of the $q$-plate, set the birefringence to $\delta=\pi/2$ by applying the correct voltage, and use circular incident polarization. In this manner, the state of the pump photons can be prepared as $\ket{\psi}_\mathrm{pump}=\pare{\ket{L,2q} + e^{i\varphi} \ket{R,0}}/\sqrt{2}$ with $q \in \{1/2,1,2/3\}$.  In contrast, in order to prepare VV pump beams, we rely on the half-wave mode of the $q$-plate by setting the birefringence to $\delta=\pi$ and using linear incident polarization. This allows us to prepare the pump photons in the state
$\ket{\psi}_\mathrm{pump} = \pare{\ket{L,2q} + e^{i\varphi} \ket{R,-2q}}/\sqrt{2}$.

The resulting vector beam is then imaged onto the dual BiBO using a telescope consisting of lenses L$_1$ ($f_1=30$ mm) and L$_2$ ($f_2=40$ mm). Each of the $500\ \mu$m thick BiBO crystals can produce SPDC photon pairs by non-collinear, frequency-degenerate, type I SPDC; note that the pump coherence length of $\sim30$m is much longer than the combined crystal length, ensuring that the contributions from the two crystals are coherently added. Note that in our experiment the pump is loosely focused at the crystal with a beam waist radius of $w_{0}$ = 2.2 mm, leading to a confocal parameter $2\pi w^{2}_{0}/\lambda$ of around 70m, many times longer than the combined crystal length (1mm).    We remark that this implies that our experiment is well within the short-crystal regime in which walkoff effects become negligible.   In this regime, note that the single photon spatial profiles and polarimetries are well inherited from the pump (with some experimental error) as is evident from figures 2 and 3. The remaining pump is suppressed by a long pass filter (transmitting $\lambda > 488$ nm), and the SPDC spectrum is limited by an $800\pm 20$ nm bandpass filter.  A lens (L$_3$) placed at a distance of one focal length ($f_3=50$ mm) from the crystals provides the Fourier transform of the crystal output Fourier plane FP$_1$. We then split the signal and the idler photons by reflecting the latter from a triangular mirror (TM). 

For the part of the experimental setup labelled as ``remote preparation", the idler photon imaged on the plane FP$_1$ is reimaged onto the Fourier plane FP$_2$ using a $4f$ telescope with 2.5$\times$ demagnification, formed by a plano-convex lens L$_4$ (with focal length $f_4=150$ mm) and a plano-convex lens L$_5$ (with $f_5=60$ mm).  Between L$_5$ and FP$_2$, the polarization of the idler photon is projected using a system (PP) consisting of either: i) a quarter-wave plate followed by a polarization beamsplitter for the $\ket{R},\ket{L}$ projections, or ii) a half-wave plate followed by a polarization beamsplitter for the $\ket{D},\ket{A}$ projections.  The tip of a multimode fiber (MMF) is placed on FP$_2$ in such a way that its 2D transverse position can be controlled by a pair of computer-controlled linear motors (25 mm travel range and 0.05 $\mu$m minimum step). The MMF, chosen to spatially filter higher modes while maintaining a relatively high coincidence counts rate ~\cite{Jabir,Vero16}, is then connected to a fiber-coupled Si avalanche photodiode (APD).

The signal photon on the plane FP$_1$ is transmitted through a quarter-wave plate (QWP$_{2}$) and a lens L$_{6}$ (with $f_{6} = 400$ mm), which directs it to the image-preserving delay line (OD) consisting of a polarizing beam splitter (PBS$_{3}$) and a series of telescopes in the $4f$ configuration, as shown in the lower right panel of figure~\ref{fig:setup}.  Note that QWP$_{2}$ and PBS$_{3}$ form a polarimeter that is used to recover the Stokes parameters of the signal photon. The delay line (OD) is composed of lenses L$_{7}$ to L$_{11}$ (with $f_{7} = f_{8}=f_{9}=750$ mm and $f_{10} = f_{11} = 100$ cm). A quarter-wave plate (QWP$_3$) with its fast axis oriented at 45 degrees is used to ensure that upon its return trip (following reflection at M$_3$) the signal photon is  deterministically reflected by PBS$_{3}$, thus exiting the delay line towards the detection stage. Note that the signal photon on plane FP$_{1}$ is relayed (via L$_6$ and L$_7$) onto a plane FP$_{1(1)}$ on M$_3$, which is then relayed (via L$_8$ and L$_9$) onto plane FP$_{1(2)}$ on M$_5$, and subsequently (via L$_{10}$ and L$_{11}$) onto plane FP$_{1(3)}$ on M$_7$.  On its return trip, the signal photon is relayed back to FP$_{1(2)}$, then onwards back to FP$_{1(1)}$, from which it is relayed (via L$_7$ and L$_{12}$, the latter with focal length $f_{12}= 400$ mm).
The plane FP$_1$ is relayed to the plane FP$_3$, which coincides with the position of the ICCD sensor, by means of the lens L$_{12}$ with focal length $f_{12}= 400$ mm.  The total delay provided by OD is 65 ns, which is sufficient to slightly overcompensate for the insertion delay of the ICCD camera.

The detection of the idler photon, projected to its zero OAM contribution and to one of the four polarizations $\vert H \rangle_i$, $\vert V \rangle_i$, $\vert L \rangle_i$, and $\vert R \rangle_i$, is used to trigger the detection of the signal photon by the ICCD camera after passing through the spatially-resolved polarimeter. In such a polarimeter, the transmission is modulated by the orientation angle of QWP$_{2}$ according to the following expression:
\begin{equation}
I(\theta) = S^{\prime}_{0} = \mathbf{M} \cdot \mathbf{S} = \dfrac{1}{2}\pare{S_{0} + S_{1}\cos^{2}{2\theta} + S_{2}\sin{2\theta}\cos{2\theta}-S_{3}\sin{2\theta}},
\label{Eq:Polarimeter}
\end{equation}
where $\mathbf{S}$ is the Stokes vector of the analyzed light, $S^{\prime}_{0}$ is the intensity after the polarimeter, and $\mathbf{M}$ is the polarimeter response, which is a function of the orientation angle of QWP$_{2}$. We rotate QWP$_{2}$ in  $\pi/8$ radian steps, collecting a total of 8 transverse intensity patterns. According to  equation \ref{Eq:Polarimeter} this system is described by $S^{\prime}_{0} = \mathbf{M} \cdot \mathbf{S}$, so that using the pseudo-inverse of $\mathbf{M}$ we get $\mathbf{M^{-1}}  \ S^{\prime}_{0} =\mathbf{M^{-1}} \cdot \mathbf{S} = \mathbf{S}$, which leaves us with $\mathbf{S}$, from which we can plot the polarization ellipses in the transverse profile of the photon~\cite{Chipman}.
\begin{figure*}[h]
    \centering
    \includegraphics[width = 0.8\textwidth]{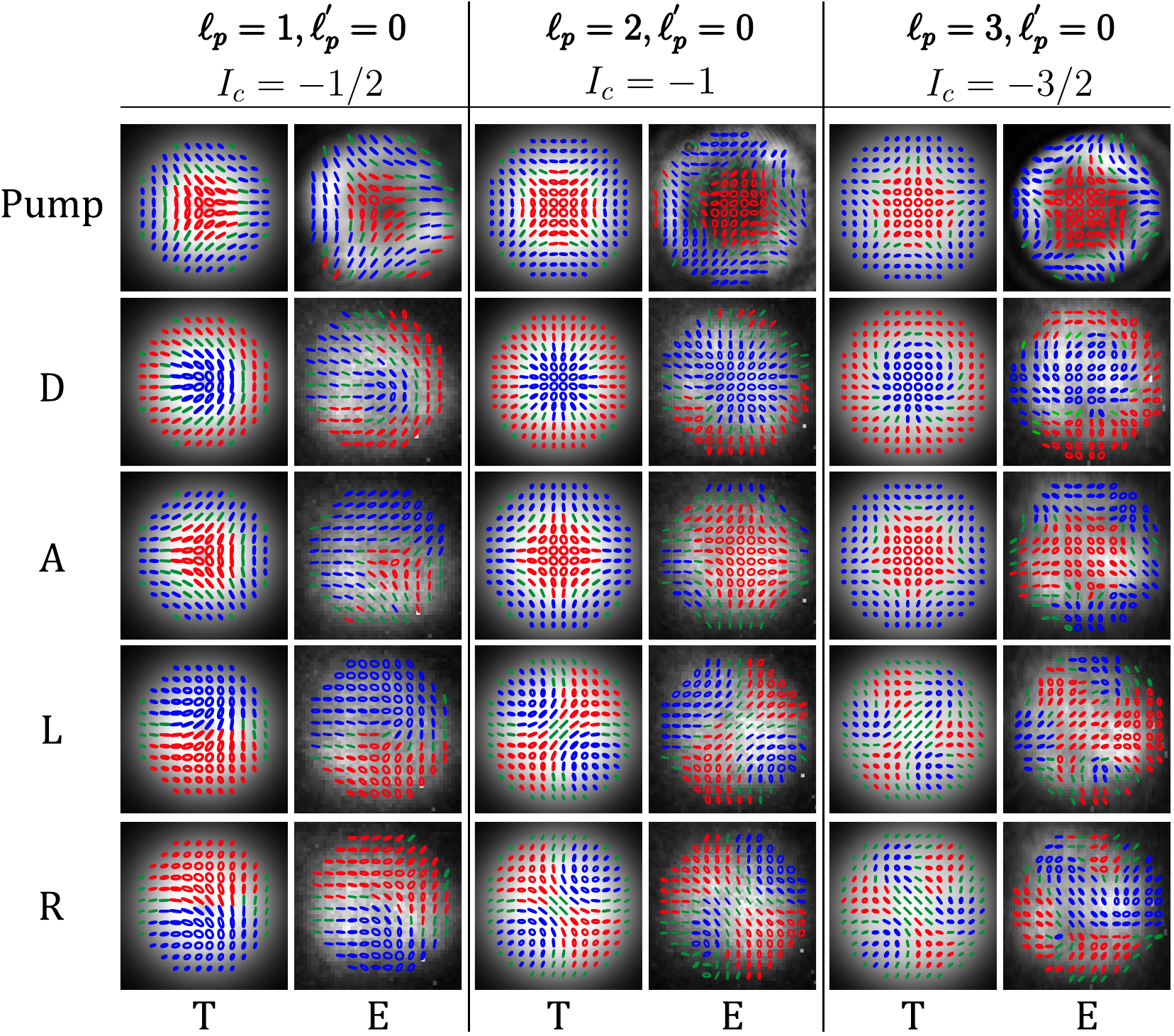}
    \caption{Vector beam structure for the pump and the heralded signal photon in the case of quarter-waveplate $q$-plate operation leading to a FP mode profile 
    (the transverse intensity shown in grayscale and the position-dependent polarization in a tricolor map, where green indicates linear polarization and blue (red)  indicates left (right) polarization handedness).  While the first row corresponds to the pump, the four remaining rows correspond to the heralded signal photon when the idler is projected to each of the zero-OAM states $\vert D \rangle$, $\vert A \rangle$, $\vert L \rangle$, $\vert R \rangle$.   The columns are organized in pairs, where the left-hand column represents a simulation (T) while the right-hand column represents the corresponding experimental measurement (E).}
    \label{fig:FP_Todos}
\end{figure*}

To ensure that both crystals contribute coherently to the two-photon state, an auxiliary polarization correlation measurement with standard coincidence detection via two avalanche photodiodes in front of the dual crystals was performed~\cite{Kwiat1999} . For this purpose, a diagonally polarized  Gaussian beam  was used to generate the SPDC pair, and with a fixed polarization projection of the idler photon and a variable polarization projection of the signal photon, the polarization correlation was recovered as shown in Fig. ~\ref{fig:setup} (bottom left), from which we obtain visibilities of 98.12\% for the idler projection to H or V and 96.17\% for the idler projection to A or D. As is well known, a unit visibility over different bases is consistent with maximal polarization entanglement.
\begin{figure*}[h]
    \centering
    \includegraphics[width = 0.8\textwidth]{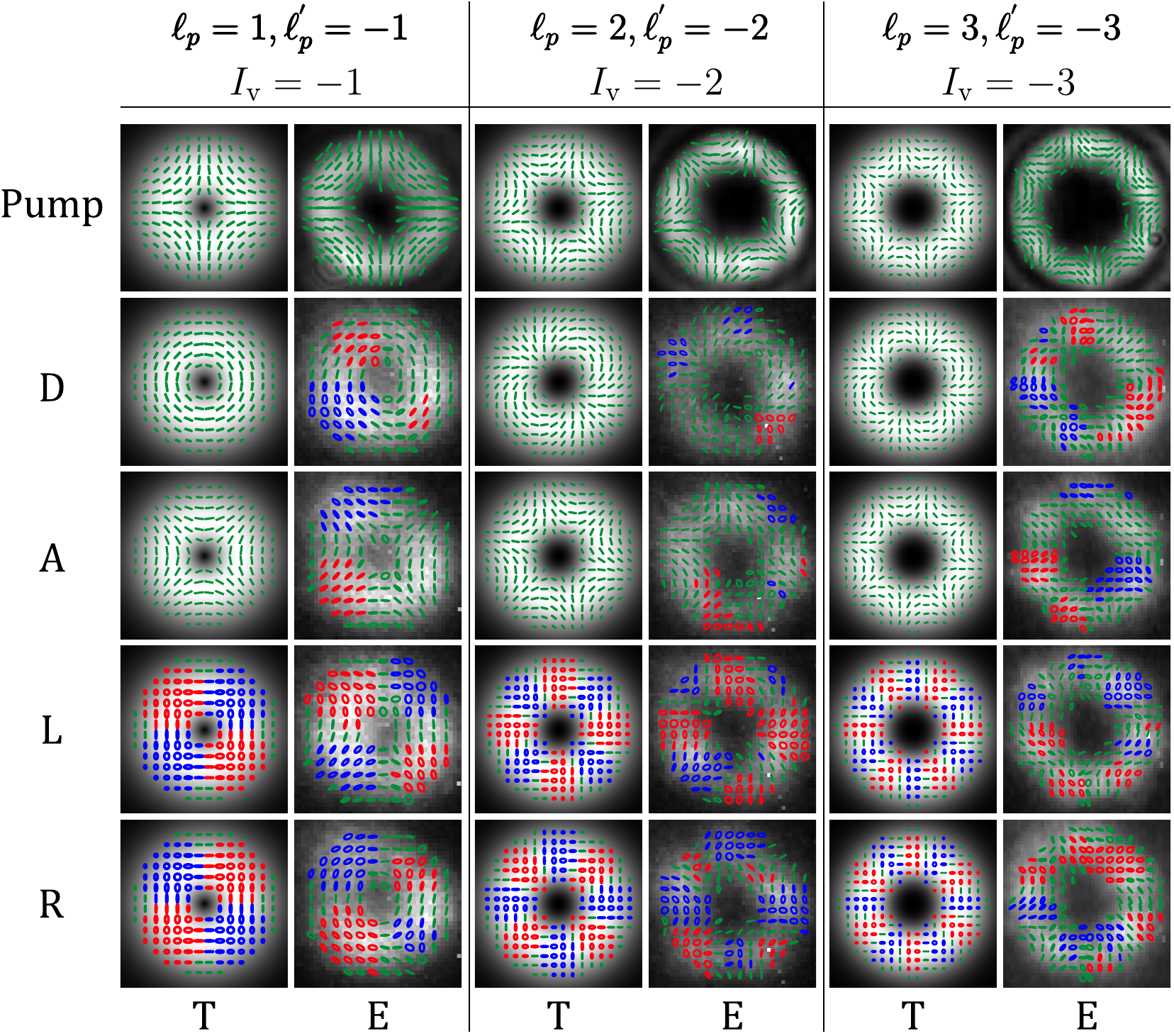}
    \caption{Vector beam structure for the pump and the heralded signal photon in the case of half-wave plate $q$-plate operation leading to a VV mode profile (the transverse intensity is shown in gray scale and the position-dependent polarization in a tricolor map, where green indicates linear polarization and blue (red) indicates left (right) polarization handedness).  While the first row corresponds to the pump, the remaining four rows correspond to the signal photon when the idler is projected into each of the zero-OAM states $\vert D \rangle$, $\vert A \rangle$, $\vert L \rangle$, $\vert R \rangle$.   The columns are arranged in pairs, with the left column representing a simulation and the right column representing the corresponding experimental measurement.
}
\label{fig:Cil_Todos}
\end{figure*}
\section*{Discussion}
The main results of this work are shown in Figures~\ref{fig:FP_Todos} and \ref{fig:Cil_Todos}, for the specific cases of Full Poincar\'e beams and Vector Vortex beams, respectively. Since we have used three different $q$-plate devices with $q$ values: $1/2$, $1$, and $3/2$, each of these two figures is organized into three groups of two columns labeled according to the value of $q$. Within each group, the left-hand column shows a simulation of the transverse intensity and position-dependent polarimetry, while the right-hand column shows the corresponding experimental measurement. In each panel, the transverse intensity is plotted in grayscale, while the polarization ellipses are plotted on a matrix of points according to a tricolor map, where green represents linear polarization and blue (red) represents left (right) polarization handedness. Note that while the first row corresponds to a measurement of the pump state produced by each of the three $q$ plates, each of the remaining four rows corresponds to the transverse intensity/position dependent polarimetry for the signal photon as the idler photon is projected into each of the following zero OAM states: $\vert D,0 \rangle_i$, $\vert A,0 \rangle_i$, $\vert R,0 \rangle_i$, or $\vert L,0 \rangle_i$, as indicated.  
The pump vector polarization prepared in the case of FP beams (quarter-wave $q$-plate operation) as well as the VV beams (half-wave $q$-plate operation) are shown in Table~\ref{Tab:PreparedPump} for each of the three $q$-plate devices. For the sake of clarity we have omitted the normalization factor $1/\sqrt{2}$ in this  quantum-inspired ket notation.
\begin{table}[h!]
	\centering
	\begin{tabular}{>{\centering\arraybackslash}m{1.5cm} >{\centering\arraybackslash}m{3cm}
			>{\centering\arraybackslash}m{3cm}
			>{\centering\arraybackslash}m{3cm}
			>{\centering\arraybackslash}m{3cm}}
        \multicolumn{3}{c}{\textbf{Prepared pump}} \\
		\hline 
		q & FP beams & VV beam \\
		\hline
		\hline
		1/2   & $\pare{\ket{L,1} + e^{i\varphi} \ket{R,0}}$ & $\pare{\ket{L,1} + e^{i\varphi} \ket{R,-1}}$\\ 
		1 & $\pare{\ket{L,2} + e^{i\varphi} \ket{R,0}}$ & $ \pare{\ket{L,2} + e^{i\varphi} \ket{R,-2}}$   \\ 
		3/2  & $\pare{\ket{L,3} + e^{i\varphi} \ket{R,0}}$ & $\pare{\ket{L,3} + e^{i\varphi} \ket{R,-3}}$ \\ 
		\hline
	\end{tabular} 
	\caption{Resulting vector polarized pump for each $q$-plate device in quarter-wave operation to prepare FP beams and in half-wave operation to prepare VV beams. We have omitted the normalization factor $1/\sqrt{2}$ for the sake of clarity while the $\varphi$ value is displayed in the main text.}
	\label{Tab:PreparedPump}
\end{table}
Note that we adjust the value of $\varphi$ in the six cases so that the transverse intensity / position-dependent polarimetry matches the azimuthal orientation of the experimental measurement; we thus obtain $\varphi=0$ for $q=1/2$, $\varphi=\pi$ for $q=1$, and $\varphi=\pi/4$ for $q=3/2$.
As expected (see theory section above), when we detect the idler photon following its projection to each of the $\vert D,0 \rangle_i$, $\vert A,0 \rangle_i$, $\vert R,0 \rangle_i$ and $\vert L,0 \rangle_i$ states, the resulting heralded signal photon is characterized by a vector structure which either corresponds directly to that of the pump (except for a rigid rotation), for projection to $\vert A,0\rangle_i$, or is the result of one of three transformations: i) the roles of the $R$/$L$ polarizations is permuted, ii) the basis is changed according to $\{L,R\}\rightarrow \{A,D\}$, or iii) the basis is changed in addition to then permuting the roles of the $A$/$D$ polarizations.  
We thus extend the ability to transfer the transverse intensity and phase from the pump to a heralded single photon from an SPDC photon pair\cite{Vero16,Sibeiro,Jabir}, to in addition also transferring the full vector polarization structure.   Interestingly, by also controlling the polarization to which the idler photon is projected, we are able to exert useful control over the properties of the heralded signal photon.    This control includes the possibility of modifying the topological class, in particular the characteristic disclination index $I_{C}$ to which the single photon vector structure belongs;  for example, the FP with $q=1/2$ case produces first-order polarization singularities, with disclination index $I_{C}=\pm 1/2$, and we can transition from a lemon structure when projecting to $\vert D,0\rangle_i$ to a star structure when projecting to $\vert A,0\rangle_i$, to a bipolar structure when projecting to $\vert L,0\rangle_i$ or $\vert R,0\rangle_i$. This topological class transition also occurs for the higher-order disclinations with $I_{C}= \pm 3/2$, produced when $q=3/2$, and transitioning from a hyperlemon structure when projecting to $\vert D,0\rangle_i$ to a hyperstar structure when projecting to $\vert A,0\rangle_i$), to a hexapolar structure when projecting to $\vert L,0\rangle_i$ or $\vert R,0\rangle_i$. The structures produced by $q=1$ transition from a radial FP beam when projecting to $\vert D,0\rangle_i$ to a hyperstar structure with $I_{C}=-1$ when projecting to $\vert A,0\rangle_i$), to a quadrupolar structure when projecting to $\vert L,0\rangle_i$ or $\vert R,0\rangle_i$ . Thus, beyond pump-single heralded photon amplitude transfer,  we are able by projecting and detecting the idler photon to, non-locally,  determine the vector polarization structure of the signal photon which in some cases presents slight variations from the expected pattern. These discrepancies, e.g. the appearance of elliptical or circular polarization components in cases where linear polarization is expected (see projections A and D in Fig~\ref{fig:Cil_Todos}) could be due to a misalignment introduced by the QWP rotation in the context of a slight non-concentricity of the two overlapped modes that make up the VV or FP modes at the single photon level; such imperfections can be significant considering that the LG mode diameter is in the order of hundreds of microns. 
%In Figure~\ref{fig:ConcentricModes}, we show a simulation of the transverse intensity and of the polarization distribution resulting from making the two constituent modes slightly non-concentric, in comparison to the ideal case.  Note that while the pair of modes being superposed is identical in both scenarios, the effect of the lateral relative shift of the constituent modes on the polarization distribution clearly resembles the behavior observed in our experimental measurements. This effect could apply to the two pump constituent modes (to be ‘inherited’ by the signal and idler photon pair) or could apply to the two coherently-added contributions to the two-photon state from the two SPDC crystals in our source.
%
\begin{figure}[t]
    \centering
    \includegraphics[width = 0.4\textwidth]{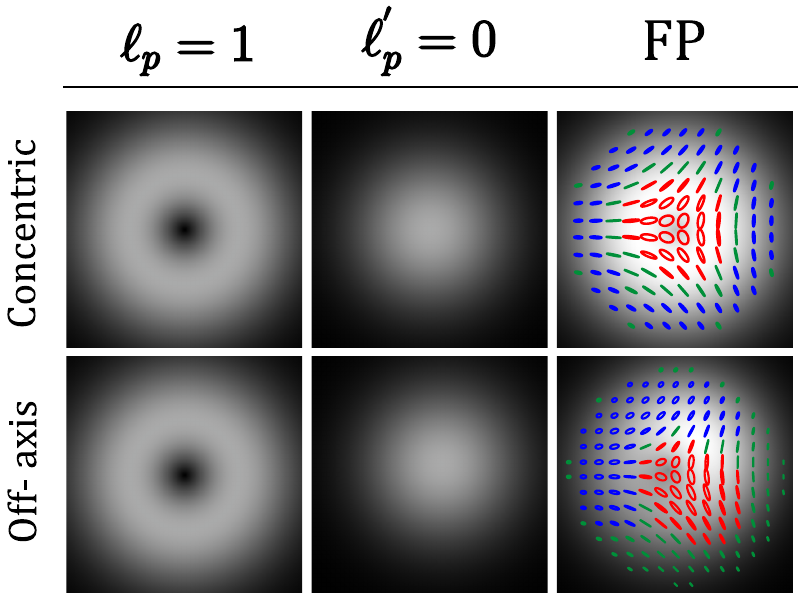}
    \caption{This figure illustrates the effect of imperfect overlap between the two  constituent LP modes of an FP mode.    In the upper row we show the ideal case involving two concentric modes giving rise to an ideal FP beam (third column).  The bottom row shows the effect of a slight  non-concentricty between the two constituent LP modes, leading to a noticeable polarization distribution deformation. The colormap is as the shown in Figures~\ref{fig:FP_Todos} and \ref{fig:Cil_Todos}}.
    \label{fig:ConcentricModes}
\end{figure}
\begin{figure}[h!]
    \centering
    \includegraphics[width = 0.5\textwidth]{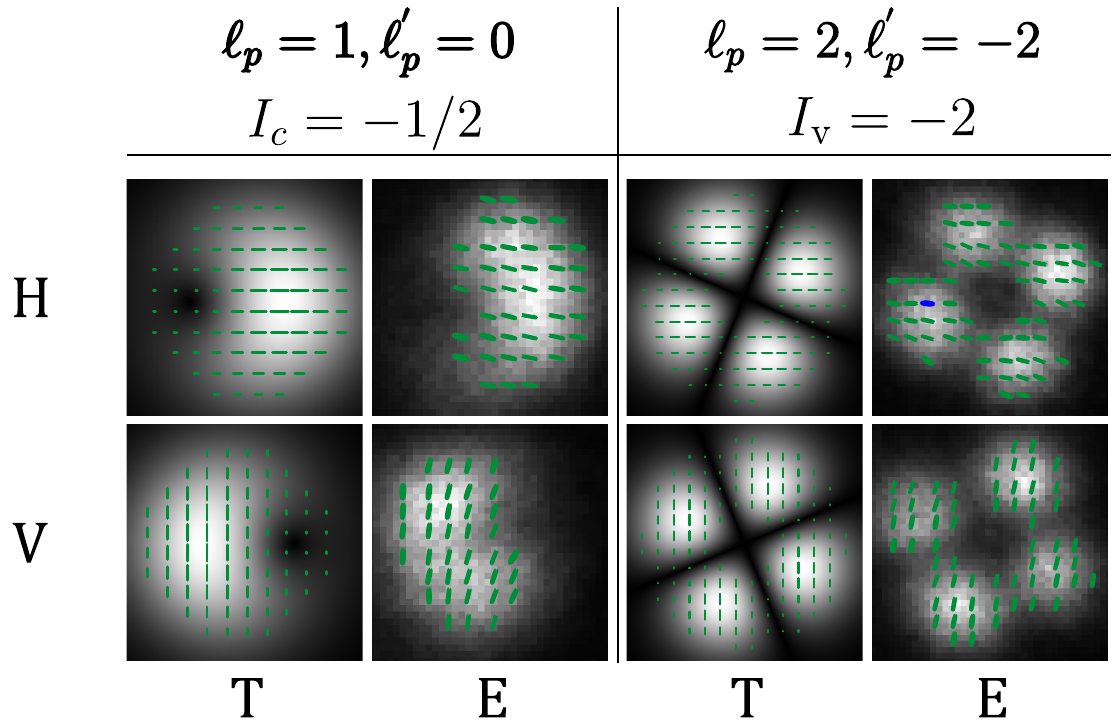}
    \caption{The effect of projecting to the state $\vert H,0\rangle$ (first row) or  $\vert V,0\rangle$ (second row) is shown, for comparison to Figure \ref{fig:FP_Todos}  (case $q=1/2$) and to Figure \ref{fig:Cil_Todos} (case $q=1$). The columns are organized in two groups; while the first column corresponds to a $q=1/2$ $q$-plate in which the pump was prepared in the state $\pare{\ket{L,1} + e^{i\varphi} \ket{R,0}}/\sqrt{2}$,  the second one corresponds to $q=1$ $q$-plate with a pump state given by $\pare{\ket{L,2q} + e^{i\varphi} \ket{R,-2q}}/\sqrt{2}$.   Within each group: the left column shows a simulation (T) and the right column shows the corresponding experimental measurement (E). As can be seen, all the polarization ellipses are in fact linear according to the tricolor map stated in Figures~\ref{fig:FP_Todos} and \ref{fig:Cil_Todos}}. 
    \label{fig:HV_Proj}
\end{figure}
In Figure~\ref{fig:ConcentricModes}, we show a simulation of the transverse intensity and of the polarization distribution resulting from making the two constituent modes slightly non-concentric, in comparison to the ideal case.  Note that while the pair of modes being superposed is identical in both scenarios, the effect of the lateral relative shift of the constituent modes on the polarization distribution clearly resembles the behavior observed in our experimental measurements. This effect could apply to the two pump constituent modes (to be ‘inherited’ by the signal and idler photon pair) or could apply to the two coherently-added contributions to the two-photon state from the two SPDC crystals in our source.
As has been pointed out, the correct pump-single heralded photon full vector amplitude transfer requires the use of a dual crystal so that both the horizontally and vertically polarized components are individually transferred.    Let us briefly consider examples of experimental measurements in which the effect of one of the two crystals is nullified, obtained by projecting the idler photon to either $\vert H,0 \rangle_i$ or $\vert V,0 \rangle_i$.  In figure \ref{fig:HV_Proj}, we present measurements similar to those in the previous two figures, obtained for a FP pump beam with $q=1/2$ (first and second columns), and for a VV pump beam with $q=1$ (third and fourth columns). In this figure, while the first row corresponds to projecting on $\vert H,0 \rangle_i$ the second row corresponds to projecting on $\vert V,0 \rangle_i$.  The important point to notice is that in all of these measurements, the measured polarization pattern is homogeneous, i.e. the vector polarization structure is suppressed.  Furthermore, it may be seen that there is no indication of pump-single heralded photon transfer, as there was in the two previous figures.
\section*{Conclusions}
In this paper, we have demonstrated an experimental protocol for the heralding of single photons described by a controllable vector polarization structure. Our work relies on the use of voltage-controlled $q$-plate devices, for the robust preparation of Vector Vortex (VV) and Full Poincare (FP) beams.  We remark that our $q$-plate approach eliminates the need for a free-space interferometer, thus yielding a simplified and more stable experimental setup. Such high-quality vector beams are then used as pump in a spontaneous parametric downconversion (SPDC) photon-pair source.  We have, for the first time to our knowledge, demonstrated the transfer of the full amplitude of a vector beam (including intensity, phase, and vector polarization) to a  single photon, heralded from the SPDC photon pair by projection and detection of its idler twin.  Furthermore, we have shown that we can control, non-locally, the resulting signal-photon vector amplitude by selecting the particular projection of the idler photon,  to one of the following zero-OAM states $\vert A,0 \rangle_i$, $\vert D,0 \rangle_i$, $\vert L,0 \rangle_i$, and $\vert R,0 \rangle_i$.  This control enables us,  on the one hand, to obtain the heralded single photon described by a pump-transferred amplitude in the $\{L,R\}$ or $\{A,D\}$ bases, or in versions of these with permuted roles for $R$/$L$ and $A$/$D$,  and on the other hand it allows us to, again non-locally, to modify at will the topological class of the resulting heralded vector structure.  We expect that this type of control over the vector structure of single photons will open up exciting new possibilities for quantum communications and information processing. 

\section*{Acknowledgments}
This work was funded by the Consejo Nacional de Ciencia y Tecnolog\'{i}a (CF-2019-217559); PAPIIT-UNAM (IN103521) and AFOSR (FA9550-21-1-0147); B.P. was supported by PNRR MUR project PE0000023-NQSTI; V.V.-H thanks Department of Physics "E. Pancini" UniNA. S.C-A. thanks CONAHCYT for the scholarship given during the development of this project.

\section*{Author contributions statement}
S.C.-A. and Z.I.-B contributed equally to this work. S.C.-A., Z.I.-B, O.C.-L, T,M.-Q and H.C.-R. carried out the experiment. D.L.-M, S.C.-A and Z.I.-B carried out the data analysis. B.P. and V.V.-H designed and manufactured the q-plates. A.B.U., S.C.-A, Z.I.-B, O.C.-L and D.L.-M developed the main conceptual ideas and physical  interpretation. S.C.-A., O.C.-L., Z.I.-B, B.P. and V.V.-H lead in writing the manuscript. A.B.U contributed to analysis and manuscript writing. A.B.U. directed and supported the project.
\bibliographystyle{ieeetr}
\bibliography{VectorBeams_bib}

\end{document}